\documentclass[sigconf, 10pt]{acmart}
\usepackage{graphicx}
\usepackage{tabularx} % for adjustable-width columns
\usepackage{xspace}
\usepackage{subcaption}
\usepackage{cleveref}
\usepackage{enumitem}
\usepackage{soul}
\AtBeginDocument{%
  }
\usepackage{tikz}
\newcommand*\emptycirc[1][.8ex]{\tikz\draw (0,0) circle (#1);} 
\newcommand*\halfcirc[1][.8ex]{%
  \begin{tikzpicture}
  \draw[fill] (0,0)-- (90:#1) arc (90:270:#1) -- cycle ;
  \draw (0,0) circle (#1);
  \end{tikzpicture}}
\newcommand*\fullcirc[1][.8ex]{\tikz\fill (0,0) circle (#1);}

\acmYear{2024}\copyrightyear{2024}
\acmConference[Preprint]{Accepted in MobiCom 2024}{Accepted in MobiCom 2024}{}
\acmBooktitle{Preprint, Accepted for Publication at MobiCom 2024}
\acmDOI{10.1145/3636534.3690684}

\author{Hossein Khalili}
\affiliation{%
  \institution{University of California, Los Angeles (UCLA)}
  \city{Los Angeles}
  \state{CA}
  \country{USA}
}

\author{Seongbin Park}
\affiliation{%
  \institution{University of California, Los Angeles (UCLA)}
  \city{Los Angeles}
  \state{CA}
  \country{USA}
}

\author{Vincent Li}
\affiliation{%
  \institution{University of California, Los Angeles (UCLA)}
  \city{Los Angeles}
  \state{CA}
  \country{USA}
}

\author{Brandan Bright}
\affiliation{%
  \institution{University of California, Los Angeles (UCLA)}
  \city{Los Angeles}
  \state{CA}
  \country{USA}
}

\author{Ali Payani}
\affiliation{%
  \institution{Cisco Systems}
  \city{San Jose}
  \state{CA}
  \country{USA}
}

\author{Ramana Rao Kompella}
\affiliation{%
  \institution{Cisco Systems}
  \city{San Jose}
  \state{CA}
  \country{USA}
}

\author{Nader Sehatbakhsh}
\affiliation{%
  \institution{University of California, Los Angeles (UCLA)}
  \city{Los Angeles}
  \state{CA}
  \country{USA}
}

\renewcommand{\shortauthors}{Khalili et al.}

\renewcommand\footnotetextcopyrightpermission[1]{} % Removes footnote with conference info

\begin{document}

\title{LightPure: Realtime Adversarial Image Purification for Mobile Devices Using Diffusion Models}\thanks{This is a preprint of a paper accepted for publication at MobiCom 2024. The final version will be available in the ACM Digital Library.}

\newcommand{\sys}{LightPure\xspace}

\begin{abstract}
  Autonomous mobile systems increasingly rely on deep neural networks for perception and decision-making. While effective, these systems are vulnerable to adversarial machine learning attacks where small perturbations in the input could significantly impact the outcome of the system. Common countermeasures include leveraging adversarial training and/or data or network transformation. Although widely used, the main drawback of these countermeasures is that they require full and invasive access to the classifiers, which are typically proprietary. Additionally, the cost of training or retraining is often prohibitively expensive for large models. To tackle this, purification models have recently been proposed. The aim is to incorporate a ``purification'' layer before classification, thereby eliminating the necessity to modify the classifier. Despite their effectiveness, state-of-the-art purification methods are compute-intensive, rendering them unsuitable for mobile systems where resources are constrained and large latency is not desired. 

This paper presents a new approach, \sys, that enhances the purification of adversarial images. It improves the accuracy of the current leading purification methods while also providing notable enhancements in speed and computational efficiency, making it suitable for mobile devices with limited resources. Our approach uses a two-step diffusion and one-shot Generative Adversarial Network (GAN) framework for purification, prioritizing latency without compromising robustness. We propose several new techniques in designing our model to achieve a reasonable balance between classification accuracy and adversarial robustness while maintaining a desired latency. We design and implement a proof-of-concept on a Jetson Nano board and evaluate our method using several attack scenarios and datasets. Our results show that \sys can outperform existing purification methods by up to 10x in terms of latency while achieving higher accuracy and robustness for various black-, gray-, and white-box attack scenarios. 
The fusion of speed and robust defense mechanisms positions our method as a significant advancement in the field of adversarial image purification, offering a scalable and effective solution for real-world mobile systems.
\end{abstract}





\maketitle

\section{Introduction} \label{sec:intro}
% Importance of machine learning based on real-time mobile applications. 
% Adversarial machine learning and its impact. 
% Defense solutions and their challenges
% purification and its challenges, our solution
% Figure 1 description
% advantages of our approach
% experiments and contributions

Deep learning models have been increasingly integrated into various aspects of decision-making in embedded and mobile devices. For example, deep learning-based computer vision techniques are now commonly used in commercial off-the-shelf advanced autonomous mobile systems (e.g., cars, robots, drones)~\cite{zhu2021adversarial,zhang2022towards,zhu2023tilemask,liu2020computing,muhammad2020deep,floreano2015science,sie2023batmobility}.

Despite their exceptional performance in various machine learning tasks, deep neural networks are vulnerable to adversarial attacks~\cite{goodfellow2014explaining,szegedy2014intriguing,carlini2017towards} where small perturbations in the input could greatly alter the output of the classifier. In the context of real autonomous mobile systems, such errors could lead to catastrophic failures and critical damages~\cite{cao2019adversarial,cao2021invisible,checkoway2011comprehensive,humayed2017cyber,liu2019edge}.  

Developing new attack and defense mechanisms for adversarial machine learning attacks has been an active area of research in the past few years~\cite{bai2021recent} and many different attack and defense strategies have been proposed. On the defense side, the main solutions can be categorized into \textit{adversarial training} \cite{dolatabadi2022l, jia2022adversarial}, \textit{data transformation} \cite{bhagoji2018transformations}, \textit{gradient masking} \cite{papernotpractical}, and \textit{moving target defenses} \cite{song2019moving}. Despite their many advantages, these methods often have poor generalization ~\cite{laidlaw2021perceptual,dolatabadi2022l,nie2022diffusion}. Additionally, adversarial training tends to be more computationally complex than standard training \cite{wong2019fast}. This complexity increases further when continuous re-training of the classifier is necessary \cite{song2019moving}. Moreover, these methods require full access to the classifier, which is often proprietary. This makes them impractical for real-world autonomous systems due to the large models, limited computational power on the mobile device, and the need for retraining the classifier. For instance, the state-of-the-art moving target defense (MTD) technique \cite{song2019moving} demands continuous training and additional models, resulting in significant overheads such as area, latency, and power consumption.

%Furthermore, the computational complexity of adversarial training is usually higher than standard training~\cite{wong2019fast}, especially when continuous re-training is needed for the classifier. Lastly, these methods require invasive and full access to the classifier which are often proprietary~\cite{jiang2022primask,lee2019occlumency,sun2023shadownet}. These solutions are not practical for real-world autonomous mobile systems as their models are often large, and private and the computational power on the device is limited. For example, the state-of-the-art \textit{moving target} defense (MTD)~\cite{song2019moving} requires \textit{continuous training} and 3-20 \textit{additional models} to achieve security, \textbf{\textit{imposing significant overheads}} (area, latency, and power overhead). 

To address these issues, new methods based on \textit{adversarial purification} have been proposed \cite{nie2022diffusion}. The key idea is to develop a machine learning block to purify adversarially perturbed images \textit{before} classification. In comparison to adversarial training methods, purification can defend against unseen threats in a cascadable plug-and-play manner without retraining the classifiers. In the context of autonomous mobile systems, such a defense strategy is much more desirable because retraining the classifier is either not feasible (proprietary) and/or extremely time-consuming. Furthermore, purification methods do not make assumptions on the form of attack and the classification model and thus can defend different pre-existing classifiers, making them more applicable in diverse scenarios (e.g., different models of autonomous systems, applicability in various geographical locations, etc.).

Early attempts at purification primarily concentrated on encoder-based and GAN-based techniques for data transformation~\cite{defensegan,meng2017magnet}. However, they proved inadequate in ensuring robustness against powerful and adaptable attacks, such as white-box projected gradient descent~\cite{nie2022diffusion,zhang2023diffsmooth}. Recently, more potent purification models have emerged, utilizing \textbf{\textit{diffusion models}}~\cite{sohl2015deep}. These models involve initially mixing the input with a small amount of noise through a forward diffusion process, followed by the restoration of the clean image via an iterative reverse generative process.

While more effective than existing invasive adversarial training models and earlier purification ones, the current state-of-the-art diffusion purification methods face \textbf{\textit{significant latency issues}} because both the forward and backward processes \textit{may need thousands of steps} to achieve the desired quality for the purified image. As we will show in our experiments, such a purification process \textit{takes hundreds of milliseconds} to process a single image on a standard mobile device. This makes it unsuitable for autonomous mobile systems that need low-latency processing and real-time decision-making.

To address the need for a \textit{cascadable}, \textit{generic}, \textit{non-invasive} but \textit{fast} purification model, we develop \textbf{\sys}. Our model leverages a lightweight diffusion strategy to achieve purification where the noise is added in one step and the recovery is also done in one shot. Our key idea is to develop a carefully designed GAN model instead of using an expensive iterative diffusion process. As we will explain in this paper, a naive design of the GAN, proposed in prior work~\cite{defensegan}, will result in either low robustness against adversarial and adaptive attacks, low accuracy due to poor noise recovery, and/or high latency. Instead, our design balances the robustness, accuracy, and latency by introducing several new contributions including a novel GAN structure and employing multiple new techniques during its training that combine diffusion models, GANs, and similarity-based loss estimation. 

\begin{figure}
    \centering
    \includegraphics[width=.95\columnwidth]{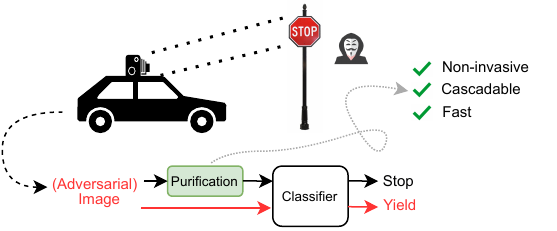}
    \vspace{-1pt}
    \caption{To protect the system from adversarial attacks, we develop a method to purify the image before classification. Our method does not require access to the classifier and is significantly faster than the state-of-the-art while maintaining robustness and accuracy.}
    \label{fig:over}
\end{figure}

A high-level overview of our method is shown in Figure \ref{fig:over}. \sys can be applied as a plug-and-play component as it does not require any prior knowledge about the classifier. Details of our design are provided in Section~\ref{sec:design}. It is also important to mention that purification is an orthogonal defense method to adversarial training, hence purification can be \textbf{\textit{combined}} with existing defense methods including MTD and adversarial training (i.e., feeding the purified images from our method to the adversarially trained classifiers). 

We develop a \textbf{\textit{proof-of-concept}} for \sys using an Nvidia Jetson Orin Nano board. Using standard datasets like CIFAR-10 and GTSRB, we show that our system can achieve strong resistance against different types of black-, gray-, and even white-box attacks, similar to leading defense methods such as adversarial training and purification. Additionally, we analyze the speed and accuracy of our model, demonstrating a significant improvement in latency compared to the best purification models for mobile systems. Our results show that \sys \textbf{\textit{outperforms existing GAN-based and diffusion-based methods}} by 2x and 10x in terms of latency, respectively, while also achieving slightly \textit{better} robustness and accuracy. Furthermore, compared to state-of-the-art non-purification methods (MTD~\cite{song2019moving}), \sys improves the latency by more than 2x while also outperforming them in terms of accuracy and robustness.  

In short, the contributions of this paper are as follows:
\begin{itemize}[leftmargin=5.5mm]

    \item A new latency-aware diffusion model based on a GAN architecture and a novel training scheme to achieve robust and low latency purification. 
    \item Improving the accuracy of the diffusion model by introducing a feedback-based accuracy-aware algorithm. 
    \item Proof-of-concept implementation of our system on a Jetson Nano embedded system.
    \item Evaluating \sys's robustness, accuracy, and latency using standard benchmarks. Comparing it with state-of-the-art adversarial defense models~\cite{song2019moving}.
    \item Our design and evaluations will be publicly available and open-sourced. 
\end{itemize}

\section{Background} \label{sec:back}
\noindent\textbf{Adversarial Attacks.} %Explain the basics of adversarial attacks. 
% I think we should explain what a good adversarial perturbation is (refer to robustbench)
Machine learning models are vulnerable to a range of attacks targeting their availability, integrity, and privacy. This paper focuses on \textit{evasion} attacks where the goal is to craft an input during inference time to cause misclassification by the victim model. More formally, the objective is to find an adversarial input, $\tilde{x}$, such that for a function, $F$, $F(\tilde{x}) \neq F(x)$, where $x$ is the original input.  

Several methods exist for crafting adversarial samples. For example, the Fast Gradient Sign Method (FGSM) was designed to attack classification models that utilize Stochastic Gradient Descent (SGD) \cite{goodfellow2014explaining}. FGSM calculates the gradients of the model's loss function with respect to each pixel, then perturbs the input data to maximize the loss function:
\begin{equation}
    \tilde{\mathbf{x}} =  \mathbf{x} + \varepsilon \operatorname{sign} \left( \nabla_{\mathbf{x}} J(\theta, \mathbf{x}, y) \right). 
\end{equation}

Given that FGSM relies on the model's parameters, it is generally considered a \textit{white-box attack}, indicating the attacker has full access to the model's architecture and parameters. 

The Projected Gradient Descent (PGD) is a more powerful multi-step variant of FGSM \cite{madry2018towards}. PGD operates on a schedule of iterative perturbations where the noise is clipped by a maximum allowed perturbation $\epsilon$ on each iteration \textit{t}. More formally:
\begin {equation}
\mathbf{x}^{(t+1)} = \text{clip}_{\mathbf{x}, \epsilon}\left(\mathbf{x}^{(t)} + \alpha \cdot \text{sign}(\nabla_\mathbf{x} J(\theta, \mathbf{x}^{(t)}, y))\right).
\end {equation}

The PGD process is solely guided by maximizing the model's loss without accounting for the confidence of the model's prediction.
 Similar to the FGSM, the PGD and its variants are also \textit{white-box attacks}. However, these attacks can still be utilized in \textit{black-box} and \textit{gray-box} scenarios.
In \textit{black-box attacks}, the internal details of the target model are completely abstracted or hidden from the attacker. Resilience to \textit{black-box attacks} is fundamental to the robustness of a model.
By training an FGSM or PGD attack on a \textit{white-box} SGD model, the attack may be effective against a \textit{black-box} target SGD model.  
A \textit{gray-box attack} lies between both \textit{white and black-box} scenarios where the attacker is given a portion of the model, while the rest is hidden. In systems with a purifier and a classifier, a possible \textit{gray-box attack} could consist of the attacker having access to the classifier but not the purifier and/or the adversary knowing the architecture of the purifier and classifier but not the internal gradients. 
In this context, defining an attacker's access to the model means the attacker knows the parameters of the model, not that they have physical access to the model.
Further details about our threat model, assumption, and attack strategies are discussed in Sections \ref{sec:design} and \ref{sec:imp}.
%moving target paper ->  https://dl.acm.org/doi/pdf/10.1145/3356250.3360025

\vspace{3pt}
\noindent\textbf{Defense Mechanisms.} Several techniques exist to defend against adversarial evasion attacks. These include adversarial training, data transformation, gradient masking, and purification. 

Adversarial training involves adding perturbed images into the training data for the classifier~\cite{croce2020reliable}. The idea is that the classifier would be able to learn what perturbations adversaries are using in attacks to improve model robustness. However, this method suffers drawbacks since generating such perturbed images for training is expensive, and this training method can decrease the classifier’s accuracy on clean images.

Data transformation involves applying linear transformations such as Principal Component Analysis (PCA) on the input data before classification \cite{bhagoji2018transformations}. The idea behind these types of defenses is that the transformations will remove the adversarial perturbations of the input before classification to improve robustness.

Gradient masking is designed to increase the difficulty of generating adversarial samples given a classifier~\cite{athalye2018obfuscated}. By purposely introducing random noise into the input of the classifier, it is harder for an adversary to determine how to perturb the input in order to degrade the performance of the classifier. This method also suffers the drawback of decreased accuracy on clean images, as purposely introducing noise can interfere with the classifier's performance. 

To address this, a \textit{moving target} defense (MTD) has recently been proposed~\cite{song2019moving,song2021deepmtd}. The key idea is to leverage multiple independently trained models during inferences to increase robustness against adversarial inputs. Furthermore, MTD dynamically retrains the networks to further improve robustness. The main drawbacks of MTD are: \textbf{\textit{(i)}} it requires multiple models (up to 20) during inferences to achieve robustness. This in turn significantly increases the latency, storage, and energy overheads. \textbf{\textit{(ii)}} MTD requires model retraining. While for small models this is possible, for large models this is not feasible. 

More recently, methods based on \textit{purification} have been proposed \cite{yoon2021adversarial, defensegan}. Purification involves using a purifying model to remove the adversarial perturbations on input images before classification The purified input is passed to the classifier for classification. The main advantage of purification is that it is independent from the classification, hence retraining the classifier is not needed. Furthermore, it is an orthogonal solution, hence it can be combined with MTD and/or adversarial training. 

\vspace{3pt}
\noindent\textbf{Adversarial Purification}. Methods based on purification are first proposed in MagNet~\cite{meng2017magnet} and DefenseGAN~\cite{defensegan}. Specifically, MagNet leverages an auto-encoder to move adversarial examples closer to the manifold of legitimate examples. The robustness is further improved by utilizing a collection of auto-encoders during runtime and having a mechanism to randomly pick one of them. 

DefenseGAN~\cite{defensegan} further improves this by introducing a GAN structure. Despite these advantages, the performance usually falls behind current adversarial training methods \cite{croce2020reliable,rebuffi2021data,jia2022adversarial}, particularly against adaptive attacks where the attacker has the full knowledge of the defense method \cite{athalye2018obfuscated,tramer2020adaptive}. This is usually attributed to the shortcomings of current generative models used as a purification model, such as mode collapse in GANs \cite{goodfellow2014generative} and the lack of randomness \cite{pinot2020randomization,nie2022diffusion}.

Recent advancements in purification research have been driven by the use of \textit{diffusion models}. This idea was first introduced in Diffpure \cite{nie2022diffusion}. It includes two steps: \textit{(i)} adding noise to the input in the forward process with small but repetitive diffusion timesteps, \textit{(ii)} removing the noise through a generative (backward) process by solving a reverse stochastic differential equation (SDE). Several follow-up works improve the speed and robustness of this process by introducing new techniques for the forward and/or backward processes \cite{nicholimproved, xiao2022DDGAN,zhang2023diffsmooth}.

More formally, given an adversarial sample $\mathbf{x}_0 \sim q(x_0)$, the first step in diffusion-based adversarial purification is adding noise to smooth the adversarial perturbations. In diffusion models \cite{ho2020denoising, sohl2015deep}, this is done via the forward diffusion process of $T$ steps and variance schedule $\{\beta_t\}_{t\in [T]}$ modeled by a conditional probability distribution: %not sure if this sequence notation is the standard... but its basically a sequence of betas from t=1 to t=T
\begin{equation}
    \begin{gathered}
        q(\mathbf{x}_{1: T} \mid \mathbf{x}_0)=\prod_{t = 1}^T q(\mathbf{x}_t \mid \mathbf{x}_{t-1}), \\
        q(\mathbf{x}_t \mid \mathbf{x}_{t-1}) =\mathcal{N}(\mathbf{x}_t ; \sqrt{1-\beta_t} \mathbf{x}_{t-1}, \beta_t \mathbf{I}).
    \end{gathered}
    \label{equ:diffusion}
\end{equation}
The total amount of noise inserted should be significant enough to remove the perturbations while preserving the global semantics of the image for accurate classification. Adversarial perturbations are typically small and therefore can be eliminated using fewer timesteps than generative tasks. Empirically, it has been shown that it is only necessary to inject noise for  7.5\% to 15\% of the total time steps required for the full diffusion process, depending on the dataset \cite{nie2022diffusion}.

%"Since adversarial perturbations are usually small, which can be removed with a small t*, the best t* in most adversarial robustness tasks also remain relatively small" (Nie et al 2022). Diffpure paper uses like t*=0.1~0.3 whereas full diffusion process is t*=1

The next step is to denoise $\mathbf{x}_T$ to obtain the purified image $\hat{\mathbf{x}}_0$, which is typically defined in diffusion models with the distribution
\begin{equation}
    \begin{gathered}
        p_\theta\left(\mathbf{x}_{0: T}\right)=p\left(\mathbf{x}_T\right) \prod_{t = 1}^T p_\theta\left(\mathbf{x}_{t-1} \mid \mathbf{x}_t\right) \\
        p_\theta\left(\mathbf{x}_{t-1} \mid  \mathbf{x}_t\right)=\mathcal{N}\left(\mathbf{x}_{t-1} ; \boldsymbol{\mu}_\theta\left(\mathbf{x}_t, t\right), \boldsymbol{\Sigma}_\theta\left(\mathbf{x}_t, t\right)\right).
    \end{gathered}
\end{equation}

However, this process is extremely slow because $T$ must be large enough to ensure that the step size $\beta_t$ is sufficiently small for the Gaussian assumption on the denoising distribution to hold \cite{feller1949theory}. This is typically in the order of hundreds. 

Our approach builds on established purification insights but introduces a new architecture tailored to address the practical drawbacks seen in current methods, especially in mobile systems. Details of our design are presented in the next section. % maybe add more here. 

%\noindent\textbf{Threat Models.} 

\section{Designing \sys} \label{sec:design}
% Overview: Diffusion models
% Our key observations: Use GAN. 
% Vanilla GAN is not satisfactory so we propose two main ideas: two-step training (with two data points and StyleGAN) and feedback-oriented adjustment (SSIM). 

The main objective of our model, \sys, is to design a fast and efficient purification model that is suitable for resource-constrained mobile systems. In the following, we first explain our threat model and assumptions and then describe the model in detail. %In this paper, we aim to enhance the latency of existing diffusion models for adversarial purification by utilizing a denoising diffusion GAN to model the denoising distribution  \cite{nie2022diffusion, xiao2022DDGAN}
% Also mention here: the adaptive attack used by diffpure authors not strong enough, so we use full gradients 
\subsection{Threat Model and Assumptions} \label{subc:threat}
We are concentrating on mobile autonomous systems that use cameras and deep neural networks for sensing and perception. Examples include autonomous cars, drones, and robots. Compared to high-end servers, these mobile devices have often limited resources including CPU and GPU computational power, memory storage, and energy. Yet, they require real-time processing given the time-sensitive cyber-physical nature of their operations (e.g., path planning, driving, etc.).  

We focus on evasion attacks during inference time, specifically image classification using a deep neural network. This is a fundamental basic block for perception in many mobile autonomous systems. We assume that the model is trained correctly and that even standard defense mechanisms such as certified robustness and adversarial training were applied during the training.

An adversary can control the inputs and can generate adversarial samples using state-of-the-art techniques. Particularly, we utilize standard and widely used AutoAttack benchmarks \cite{croce2020reliable, croce2020robustbench} for crafting adversarial samples. We consider both targeted and non-targeted attacks. Further, we assume that the adversary has either \textit{(i)} full knowledge about the internals of the system, including the structures and the internal values (gradients). This is referred to as a \textit{white-box} attack; or \textit{(ii)} partial knowledge where the attacker only has access to the gradient of the classifier and does not know anything about the purifier. We refer to this as a \textit{gray-box} attack; or \textit{(iii)} no knowledge about the system, considering the model as a \textit{black-box}.%Note that the white-box assumption is quite strong for mobile systems. Most prior research on mobile systems assumed that the attacker doesn't have physical access to the internal system values (such as gradients), hence they assumed white-box attacks out of scope. Otherwise, they could execute a more direct and potent attack without needing adversarial inputs. In this paper, we do consider all three types and report the results for all of them. We, however, emphasize that the gray-box attack scenarios are the most realistic in modern autonomous mobile systems. 

The assumption of white-box attacks is particularly strong in the context of mobile systems. Previous studies in this field operated under the assumption that attackers do not possess physical access to internal system values, such as gradients. Consequently, they considered white-box attacks to be outside the scope of their research~\cite{song2021deepmtd,song2019moving}. The rationale is that if attackers did have such access, they could potentially carry out more direct and powerful attacks without requiring adversarial inputs. This paper, however, examines all three types of attacks and presents findings for each. Nonetheless, it is highlighted that gray-box attack scenarios are the most realistic in contemporary autonomous mobile systems.

\subsection{\sys Overview}
As discussed in Section~\ref{sec:back}, purification models offer various advantages such as flexibility and suitability for autonomous mobile systems. However, the main issue is that existing purification models are slow. Our main contribution is creating a new purification model that is accurate, fast, and robust. The key insight here is that to speed up the purification process, we should \textbf{\textit{(i)}} use much larger time steps (meaning fewer iterations), and \textbf{\textit{(ii)}} make denoising (both forward and backward) less complex. Consequently, if we increase the step size, $\beta_t$ in Equation~\ref{equ:diffusion}, to decrease the number of steps, the true denoising distribution becomes more complex and multimodal. Our main observation is that a GAN model can be used since they have been proven effective in modeling such distributions in the image domain \cite{ledig2017photo, xiao2022DDGAN}. Furthermore, evaluating GANs is much more lightweight compared to SDEs which were used in prior methods, further improving the speed.

Designing a GAN-based purification model, however, is \textbf{\textit{not trivial}}. The important consideration is that in an autonomous mobile system, three important metrics should be jointly optimized: \textit{accuracy}, \textit{robustness}, and \textit{latency}. Naive design decisions could lead to poor designs where some or all metrics are sacrificed in favor of others (e.g., poor accuracy but good robustness and latency, large latency with good accuracy, etc.). To optimize this, we design \sys. Our design has two important contributions: \textbf{\textit{(i)} Latency-aware robust diffusion model} and \textbf{\textit{(ii)} Accuracy-aware training scheme}. Note that our method is fundamentally different than GAN-only methods for purification~\cite{defensegan,meng2017magnet}. These methods do not employ diffusion and as a result, are significantly less robust against adaptive attacks~\cite{nie2022diffusion}. As we will show, combining GANs in a diffusion model, however, is non-trivial and requires a careful set of considerations.

Figures \ref{fig:sys} and \ref{fig:inf} present the main steps in our design. 
In the following, we explain each component in detail.

\subsection{Latency-Aware Diffusion Model}\label{subc:denoising}
To explain our design, we first provide its high-level intuition and then present the formal details. Recall that our goal is to design a purifier that has a low number of denoising steps and each step is preferably simple. As a result, initially, we aim for a method that has only one forward step (diffusion) and one backward step (denoising). Internally, the design includes a generator that learns how to generate (denoise) an input using a discriminator that can distinguish between noisy (fake) and clean (real) images. 

\begin{figure}
    \centering
    \includegraphics[width=1\columnwidth]{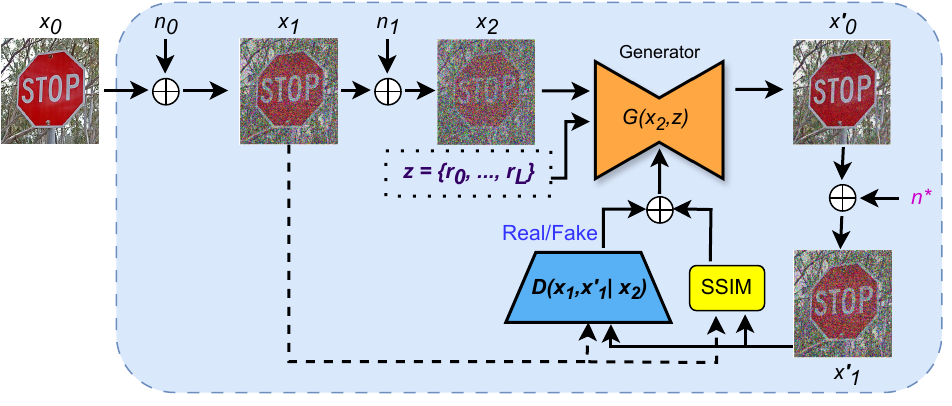}
    \caption{Training the purifier involves multiple steps. The original (clean) image is first diffused in two steps. The perturbed image is then fed into a generator. The generator is trained using a loss function that is a combination of a conditional discriminator and similarity (SSIM) losses.}
    \label{fig:sys}
\end{figure}

\begin{figure}
         \centering
         \includegraphics[width=1\columnwidth]{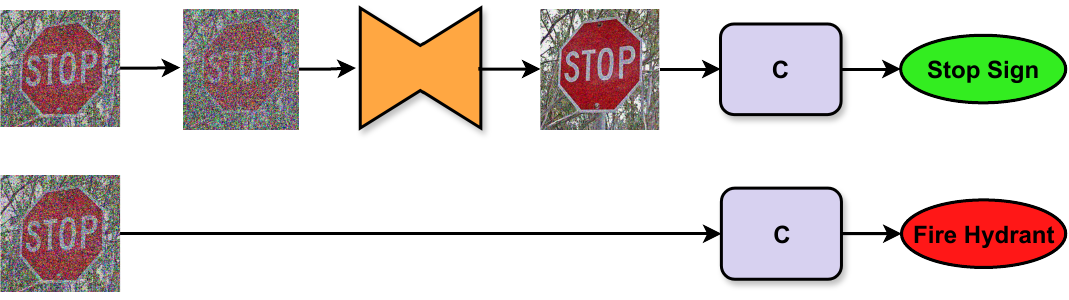}
         \caption{The trained generator is used during the inference to purify images dynamically.}
         \label{fig:inf}
\end{figure}

Our initial analysis revealed that this model lacks generalization and robustness primarily because it predominantly learns to replicate the input. While this is desirable for standard image generation tasks, it proves detrimental for adversarial purification. In such cases, closely mimicking the input is insufficiently robust, given that the input image may contain adversarial elements (recall that the objective is to \textit{eliminate} adversarial noise). The only benefit of mimicking the input is that the `clean' data accuracy remains unchanged as the purification does not significantly change the input (note that this is desired for image generation scenarios). 

To address this, we've made a new observation that a two-step diffusion (forward) process could potentially solve the issue. Essentially, our model first generates a noisy image (depicted as $x_1$ in Figure~\ref{fig:sys}) from the original clean image. Instead of using this for the discriminator and generator, we create \textit{another} noisy image by adding more noise to $x_1$ (and not the clean image). The goal is to then use these two noisy samples $(x_1, x_2)$ to train the generator and discriminator ($G()$ and $D()$ in Figure~\ref{fig:sys}). The idea is for the model to learn to denoise $x_2$ given a noisy input $x_1$. In contrast, originally (with just one step), the model learned how to denoise $x_1$ given a clean input $x_0$. The key difference lies in the fact that during inference, inputs are more similar to $x_1$ than $x_0$, hence the purification will be more robust. Note that here we don't make any assumption about the noise and/or distribution. Instead, the observation is that the model should learn to purify a noisy image rather than mimicking a clean one. In Section~\ref{sec:result}, we will show that our model remains robust under different noise distributions, further confirming this claim.  

The steps during inference are also shown in Figure~\ref{fig:inf}. Here, we only add one more step (given that the input is already noisy) and the goal is to generate a clean image from these two samples. 

To enhance robustness further, we observe that the discriminator could also benefit from a two-step approach. Instead of using $x_2$ and $G(x_2)$ for the discriminator (to distinguish between noisy and clean images), we compare $x_1$ and its noisy version, $x'_1$, which is created by adding noise to the purified image (shown as $x'_0$ in Figure~\ref{fig:sys}). The idea is for the generator function to learn how to \textit{completely} purify the image (i.e., one-shot) using $x_2$ without recreating $x_1$ (which remains noisy). In simpler terms, since the objective is to purify adversarial samples, the generator should be able to produce a clean image (similar to $x_0$) from a noisy sample, while the discriminator should be able to distinguish between an original noisy image and its fake version (purified and then perturbed again). Additionally, guiding this process using $x_2$ as a condition further improves the discriminator's accuracy (more details later).

More formally, to train our GAN for denoising with larger step sizes, we reduce the number of timesteps without having to increase $\beta_t$ by setting $T=2$ and train a time-independent discriminator and a one-shot generator. Specifically, %we employ a similar methodology used to train denoising diffusion GANs, as proposed by Xiao et al. \cite{xiao2022DDGAN}. As justified in Section \ref{sec:back}, for purification, it is not necessary to add as much noise as generative tasks; therefore, we can further reduce the number of timesteps without having to increase $\beta_t$. For our experiments, we set $T=2$ and trained a time-independent discriminator and a one-shot generator. Our training process is outlined in Figure \ref{fig:sys} (a).
from the raw image \( \mathbf{x}_0 \), we obtain \( \mathbf{x}_1 \) and \( \mathbf{x}_2 \) by diffusing for one and two timesteps, respectively:
\begin{equation}
    \mathbf{x}_1 = \sqrt{1-\beta_1}x_0 + \sqrt{\beta_1}\mathbf{\epsilon}.\label{equ:beta1}
\end{equation}
\begin{equation}
    \mathbf{x}_2 = \sqrt{1-\beta_2}x_1 + \sqrt{\beta_2}\mathbf{\epsilon}.\label{equ:beta2}
\end{equation}
where $\mathbf{\epsilon} \sim \mathcal{N}(\mathbf{0, I}_d)$ and $\beta_1, \beta_2$ is the variance schedule. 
Our generator, a one-shot model denoted as $G_\theta\left(\mathbf{x}, \mathbf{z} \right): \mathbb{R}^N \times \mathbb{R}^L \rightarrow \mathbb{R}^N$, models the denoising distribution $p_\theta (\mathbf{x}_{1} \mid \mathbf{x}_2).$ In other words, it predicts $\mathbf{x}_0^\prime$ from $\mathbf{x}_2$ and an $L$-dimensional latent variable $\mathbf{z} \sim p(\mathbf{z}):=\mathcal{N}(\mathbf{z} ; \mathbf{0}, \mathbf{I})$:
\begin{equation}
    \mathbf{x}_0^\prime = G_\theta\left(\mathbf{x_2}, \mathbf{z} \right).
\end{equation}

$\mathbf{x}_1^\prime$ is sampled using the posterior distribution $q(\mathbf{x}_{t-1}\mid\mathbf{x}_t,\mathbf{x}_0)$ given $\mathbf{x}_2$ and $\mathbf{x}_0^\prime$ (similar approach has been proposed in prior work by Xiao et al. \cite{xiao2022DDGAN}). The noise is denoted by $n^*$ in Figure~\ref{fig:sys}.

The discriminator with parameters $\phi$, denoted as \( D_\phi(\mathbf{x},\mathbf{x}_2): \mathbb{R}^N \times \mathbb{R}^N  \rightarrow [0,1]\), takes two N-dimensional inputs, \( \mathbf{x} \) and \( \mathbf{x}_2 \), and determines whether \( \mathbf{x} \) is an output from the generator ($\mathbf{x}_1^\prime$) or from a real sample ($\mathbf{x}_1$). We discriminate on a diffused input to prevent overfitting of the discriminator, since the diffusion process smooths the data distribution \cite{lyu_uai09}. The loss function of the discriminator quantifies how well the discriminator can differentiate between real and denoised samples by comparing its predictions to the actual labels of the data. It is formalized as follows:
% \begin{equation} \label{eq:4}
%     \begin{split}
%     \min _\phi \mathbb{E}_{q(\mathbf{x}_2)} &\left[\mathbb{E}_{q(\mathbf{x}_{1} \mid \mathbf{x}_2)}[-\log (D_\phi (\mathbf{x}_{1}, \mathbf{x}_2) ] \right.\\
%     & \quad \left.+\mathbb{E}_{p_\theta (\mathbf{x}_{1} \mid \mathbf{x}_2)}[-\log (1-D_\phi (\mathbf{x}_{1}, \mathbf{x}_2))] \right].
%     \end{split}
% \end{equation}
% Since the true denoising distribution \( q(\mathbf{x}_1\mid \mathbf{x}_2) \) is unknown, we must leverage the chain rule for joint probabilities to rewrite the first expectation in Equation \ref{eq:4} as:
% \begin{equation}
%     \begin{split}
%         \mathbb{E}&_{q\left(\mathbf{x}_2\right) q\left(\mathbf{x}_{1} \mid \mathbf{x}_2\right)} \left[-\log \left(D_\phi\left(\mathbf{x}_{1}, \mathbf{x}_2 \right)\right)\right]\\
%         &\qquad=\mathbb{E}_{q\left(\mathbf{x}_0\right) q\left(\mathbf{x}_{1} \mid \mathbf{x}_0\right) q\left(\mathbf{x}_2 \mid \mathbf{x}_1\right)}\left[-\log \left(D_\phi\left(\mathbf{x}_{1}, \mathbf{x}_2\right)\right)\right] .
%     \end{split}
% \end{equation}
% Then, the final discriminator loss is: 
\begin{equation} 
    \begin{split}
    \min_\phi\; &\mathbb{E}_{q\left(\mathbf{x}_0\right) q\left(\mathbf{x}_{1} \mid \mathbf{x}_0\right) q\left(\mathbf{x}_2 \mid \mathbf{x}_1\right)}\left[-\log \left(D_\phi\left(\mathbf{x}_{1}, \mathbf{x}_2\right)\right)\right]\\
    & \qquad  +\mathbb{E}_{q\left(\mathbf{x}_2\right) p_\theta \left(\mathbf{x}_{1} \mid \mathbf{x}_2\right)} \left[-\log \left(1-D_\phi\left(\mathbf{x}_{1}^\prime, \mathbf{x}_2 \right)\right)\right].
    \end{split}
\end{equation}

For the generator's architecture, we follow Xiao et al. and use StyleGAN, a non-saturating GAN variant \cite{xiao2022DDGAN, karras2019style, karras2020analyzing}. While we considered other architectures, we found that StyleGAN achieves the right balance between computational complexity and accuracy. 

Since the goal of the generator is to generate samples that the discriminator classifies as real, it is penalized by the loss function when the discriminator correctly identifies its output as fake. The loss function is formulated as follows:
\begin{equation}
    \min _\theta \mathbb{E}_{q\left(\mathbf{x}_2\right)} \mathbb{E}_{p_\theta\left(\mathbf{x}_1 \mid \mathbf{x}_2\right)}\left[-\log \left(D_\phi\left(\mathbf{x}_1^\prime, \mathbf{x}_2\right)\right)\right].
\end{equation}

Note that, we did not further try to optimize the GAN structure as our main novelty relies on the design algorithm (two-step forward and one-shot reverse training) rather than the GAN architecture itself. We left further optimization of the internal architecture to future work.
Details of the network are provided in Section~\ref{sec:imp}, and are shown in Figure~\ref{fig:gan} (see Appendix).

\subsection{Accuracy-Aware Denoising Model}\label{subc:denoising}
The objective in \sys is to optimize robustness, latency, and accuracy simultaneously. The design outlined above enhances both robustness and latency by employing a one-shot generation during inference, as depicted in Figure~\ref{fig:inf}. This approach is fast and utilizes a generalized model that is robust. However, our analysis revealed that the purification process significantly affects downstream task accuracy due to its invasive nature. To address this issue, we enhance \sys by introducing a new denoising step that considers accuracy as a factor.

Inspired by Cycle-Consistent Adversarial Networks~\cite{zhu2017unpaired}, to improve the quality of denoised images, we incorporate a metric comparing the similarity between a denoised image and the original clean image into the generator's loss function \cite{zhu2017unpaired}. Specifically, we use the Structural Similarity Index Measure (SSIM), which is based on the computation of three factors: luminance, contrast, and structure \cite{zhao2015loss, wang2004image}.  The SSIM calculates the similarity between $x_1$ and $x'_1$, adding it as an extra component to the generator's loss function. The updated loss function is 
\begin{equation}
    \min _\theta \mathbb{E}_{q\left(\mathbf{x}_2\right)} \mathbb{E}_{p_\theta\left(\mathbf{x}_1 \mid \mathbf{x}_2\right)}\left[-\log \left(D_\phi\left(\mathbf{x}_1^\prime, \mathbf{x}_2\right)\right)\right] + \lambda(1-\text{SSIM}(x,y)). \label{equ:lambda}
\end{equation}
% add sentence on why we chose ssim over other similarity measures

Here, $\lambda$ is a regularization factor, balancing the importance of accuracy vs. robustness. Details are also shown in Figure~\ref{fig:sys}. In Section~\ref{sec:result}, we will show that using SSIM as part of the loss function has a significant impact on improving the overall accuracy and robustness metrics across different datasets.

\section{Proof-of-Concept Implementation} \label{sec:imp}

%\subsection{The Nvidia Jetson Orin Nano}
\noindent \textbf{Device.} We utilize an Nvidia Jetson Orin Nano board to develop the proof of concept for \sys. This board is commonly used for running deep neural networks in various applications like automotive, manufacturing, retail, and more. It features a six-core ARM CPU, a 1024-core NVIDIA Ampere architecture GPU with 32 Tensor Cores, and 8GB LPDDR4X memory. Despite being smaller and more resource-constrained compared to other Jetson family versions, our results in Section~\ref{sec:result} demonstrate that \sys can achieve real-time latency. Furthermore, the implementation can easily apply to other versions with minimal effort. Our board uses Linux (Ubuntu 22.04) and Nvidia-provided SDK (JetPack 6.0). Pytorch 2.1.0 is used to run the codes and models with CUDA toolkit version 11.5. %\textcolor{blue}{Pytorch version?}

%The Jetson boards are commonly used in embedded  is a standard development kit for prototyping the class of mobile autonomous systems on which existing purification techniques are unsuitable. Therefore if it can be shown that a purification technique running on the Jetson can produce accurate and robust image classification in near real time speeds, then that purification technique would certainly be viable in practical applications. 

\vspace{3pt}
\noindent\textbf{Generating Adversarial Samples}
Since the computing power of the Jetson is limited, we generate the adversarial images beforehand on a separate server with access to more computing power. For generating samples and training the model, we rely on a server equipped with two Nvidia A6000 GPUs featuring NVLink, 96 GB of memory, and a 10-core Intel 11th-gen processor.  We use various datasets and attack strategies to create the baseline and adversarial samples. Details are explained in Section~\ref{sec:setup}.

\vspace{3pt}
\noindent \textbf{\sys Architecture.} The \sys purifier is developed and trained on the server (details above).  Once trained, the model weights are exported to the Jetson for testing. Our \sys purifier has 45,610,598 parameters which take up 174MB of memory. The structure of our purifier (the generator) is shown in the Appendix (Figure \ref{fig:gan}). 

Our generator structure largely follows the U-net structure~\cite{ronneberger2015u}
which consists of multiple ResNet 
and Attention blocks. As mentioned in Section~\ref{sec:design}, the key novelty in our design is the usage of a two-step forward with a one-shot reverse mechanism that was further enhanced by incorporating SSIM (similarity) loss. 

\begin{figure*}    
\begin{subfigure}[]{0.28\textwidth}
         \centering
         \includegraphics[width=\textwidth]{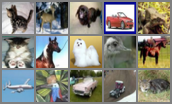}
         \caption{Clean.}
         \label{fig:clean}
     \end{subfigure}
     %\hspace{0.05\textwidth}%
\begin{subfigure}[]{0.28\textwidth}
         \centering
         \includegraphics[width=\textwidth]{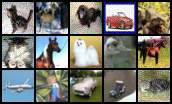}
         \caption{(Adversarial) Perturbed.}
         \label{fig:pert}
     \end{subfigure}
   %\hspace{0.05\textwidth}%
\begin{subfigure}[]{0.28\textwidth}
         \centering
         \includegraphics[width=\textwidth]{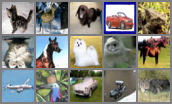}
         \caption{Purified.}
         \label{fig:black}
     \end{subfigure}
    \caption{Examples of clean images, perturbed adversarial images, and purified images on the CIFAR-10 dataset.}
    \label{fig:mo}
\end{figure*}

% \begin{figure*}
% \begin{subfigure}{0.2\textwidth}
%          \centering
%          \includegraphics[width=\textwidth]{Figures/clean_grid.png}
%          \caption{Clean.}
%          \label{fig:e3}
% \end{subfigure}
% \hspace{0.1\textwidth}%
% \begin{subfigure}{0.2\textwidth}
%          \centering
%          \includegraphics[width=\textwidth]{Figures/adv_grid.png}
%          \caption{(Adversarial) Perturbed.}
%          \label{fig:e1}
% \end{subfigure}  
% \hspace{0.1\textwidth}%
% \begin{subfigure}{0.2\textwidth}
%          \centering
%          \includegraphics[width=\textwidth]{Figures/purified_grid.png}
%          \caption{Purified.}
%          \label{fig:e2}
% \end{subfigure} 
%  \label{fig:example}
%     \caption{Examples of clean images, perturbed adversarial images, and purified images on CIFAR-10 dataset.}
% \end{figure*}

We use $\lambda =3$ for the loss function (see Equation~\ref{equ:lambda}). Also, we select 256 for the latent dimension and 512 for the latent embedding dimension ($z$). For the diffusion process, we choose $\beta_1 =0.0167$ and $\beta_2 =0.0331$ for the noise levels, and $10^{-4}$ for the learning rate.

%\textcolor{blue}{Other hyperparameters here.}

Once trained, the purifier is then paired with different classifiers (details in Section~\ref{sec:setup}) during inference. \textit{Note that, \sys does not need to know the details of the classifier nor does it require retraining the classifier}. Readers can refer to Figures \ref{fig:sys} and \ref{fig:inf} for the design and steps during the training and inference. Our code and data will be \textbf{\textit{open-source}} and publicly available.

The examples of clean, perturbed (using forward diffusion), and purified (using the generator) are shown in Figure~\ref{fig:mo}. The examples show that \sys can maintain the main visual aspects of the original image while being capable of removing the perturbations. The detailed results for accuracy and robustness are presented in Section~\ref{sec:result}.

\vspace{3pt}
\noindent \textbf{Measuring Latency.} Images are classified individually, and the time it takes to classify each image is measured. To obtain the duration for each model to return a result, we utilize Python's built-in time package's \texttt{perf\_counter()} function. We record the time just before classification starts and immediately after classification produces a result. By calculating the difference, we determine the time it took to classify one image. The latency is the averaged number over 1000 instances. This approach is employed to emphasize the actual latency, rather than measuring the end-to-end (sensor-computation-actuation) latency, which could be affected by sensor/actuator and/or network delays. Additionally, we did not take batching into account, as the assumption is that the model must process images individually in a real-world system, such as autonomous driving.

\section{Evaluation Setup} \label{sec:setup}

%For the baseline four separate sets of adversarial images were generated. The first set of adversarial images was generated with full access to the classifier used by our baseline and used APGD-CE. This is the gray box threat model. For the APGD-CE attack, we chose to use the L-infinity norm, we set the number of restarts to 1, set the number of target classes to 9, and trained with a learning rate of 8/255. These hyperparameters are mostly the same as what the creators of Diffpure themselves used to attack classifiers. The second set of adversarial images was generated with the same access to the actual classifier but with AutoAttack's random attack mode. The third and fourth sets of adversarial images are almost the same as the first two sets of adversarial images, however, the adversary only had access to a different, but similarly performing, classifier. The third and fourth sets of images represent the black box threat model. We created 1000 images in each set of adversarial images.

\vspace{3pt}
\noindent \textbf{Metrics.}
%The attack that we are using is apgd-ce it is a variation of pgd introduced by autoattack, for the final testing we need to evaluate our model with all of this attack list [apgd-ce, apgd-t, fab-t, square]
We use two main metrics, clean accuracy and robust accuracy. The `clean accuracy' or `standard accuracy' refers to the performance of the model on untampered data. When employing a purifier before the classifier, there might be a slight decrease in clean accuracy due to perturbations.  Note that this is common even for other defense mechanisms such as adversarial training. The reduction in accuracy is primarily because training on adversarial examples can inadvertently influence the model's learning in a way that slightly compromises its performance on clean data. Therefore, it is ideal to have zero or minimal impact on clean accuracy when adding purification.

We use three standard configurations for `robust accuracy'.  

\textbf{\textit{1-Black-Box Attack:}} In this setup, the adversary has no knowledge about either the classifier or the purifier. Essentially, the adversary is operating without understanding the internal mechanisms or configurations of either component.

    %2-Partial Black-Box Attack: Here, we possess information about the classifier and generate adversarial samples based on this classifier. However, we have no knowledge of the purifier. The robust accuracy is then assessed by applying the purifier to the adversarial samples to evaluate its effectiveness.

    \textbf{\textit{2-Gray-Box Attack:}} The adversary is knowledgeable about both the classifier and a purifier (their architecture), but the classifier used for generating the attack is not the one targeted in the evaluation. The attack is crafted based on this external purifier and the classifier, and the robustness is tested against the intended (targeted) purifier and classifier.

    \textbf{\textit{3-Complete White-Box Attack:}} This setup allows full access to both the classifier and the purifier including the architecture, hyperparameters, and the ability to calculate the full gradients of both the classifier and purifier. Note that in real mobile systems, this is less realistic as it requires full physical access to the device and its internal computations. Given such a strong capability, the adversary could possibly launch a more direct and more powerful attack instead of launching an ML-based evasion attack.  

\vspace{3pt}
\noindent \textbf{Attack Methods.}
To assess the robustness of our model against adversarial attacks, we employ the RobustBench benchmark, which uses AutoAttack, an ensemble of white-box and black-box attacks \cite{croce2020reliable, croce2020robustbench}. Autoattack utilizes Auto-PGD, a more powerful variant of PGD that automatically adapts the step size. This can be applied to various loss functions, including cross-entropy loss (APGD-CE) and difference of logits ratio loss (APGD-DLR):
\begin{gather}
\mathrm{CE}(x, y)=-\log p_y=-z_y+\log \left(\sum_{j=1}^K e^{z_j}\right)  \\
\mathrm{DLR}(x, y)=-\frac{z_y-\max _{i \neq y} z_i}{z_{\pi_1}-z_{\pi_3}}
\end{gather}
Due to the stochastic nature of our model, we utilize the AutoAttack version against randomized adversarial defenses (APGD-rand), which is an ensemble of APGD-CE and APGD-DLR. To counter randomness, APGD-rand applies Expectation over Transformation (EOT), which averages 20 computations of the gradient at the same point, to get the direction for the update step \cite{athalye2018obfuscated}. Additionally, we run an experiment with APGD-CE without EOT to evaluate our model against faster but less powerful attacks. Similar to prior work, we choose to use the L-infinity norm, we set the number of restarts to 1, set the number of target classes to 9, and train with a learning rate of 8/255.

\begin{figure*}    
\begin{subfigure}[]{0.32\textwidth}
         \centering
         \includegraphics[width=\textwidth]{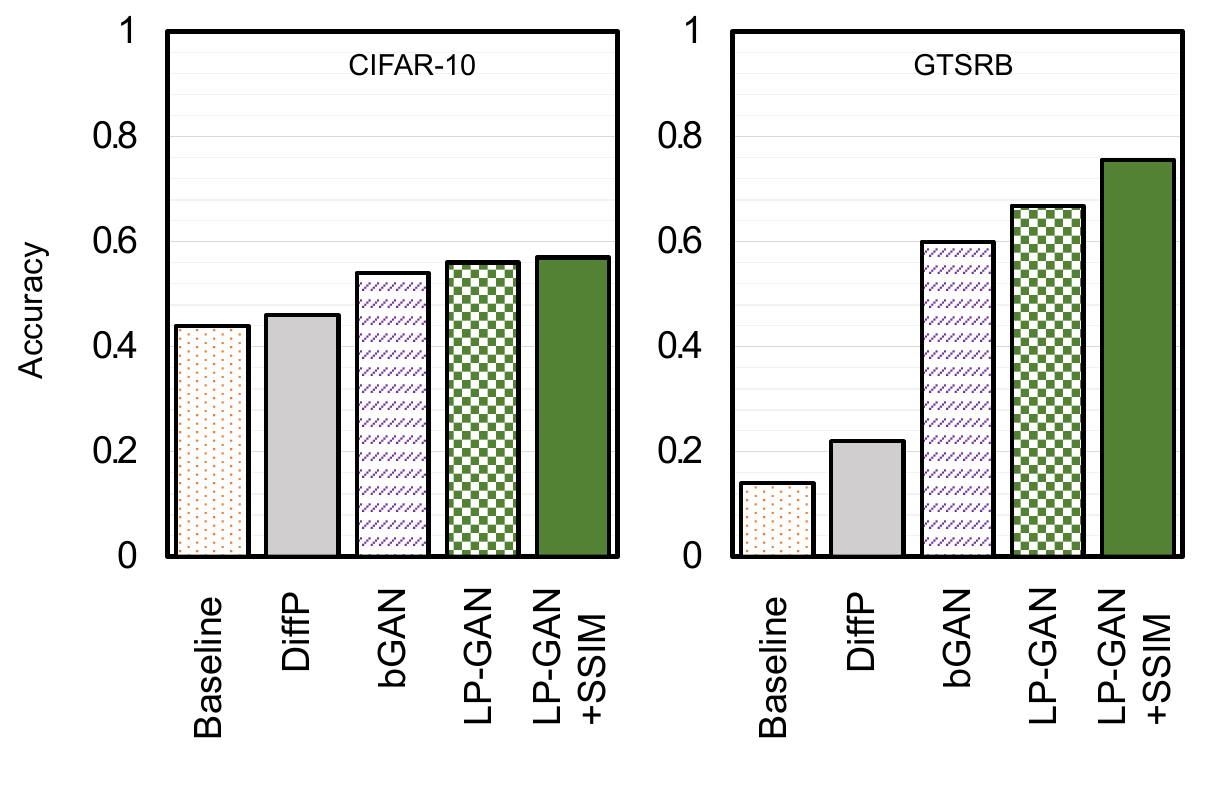}
         \caption{White-Box.}
         \label{fig:white}
     \end{subfigure}
     \hfill
\begin{subfigure}[]{0.32\textwidth}
         \centering
         \includegraphics[width=\textwidth]{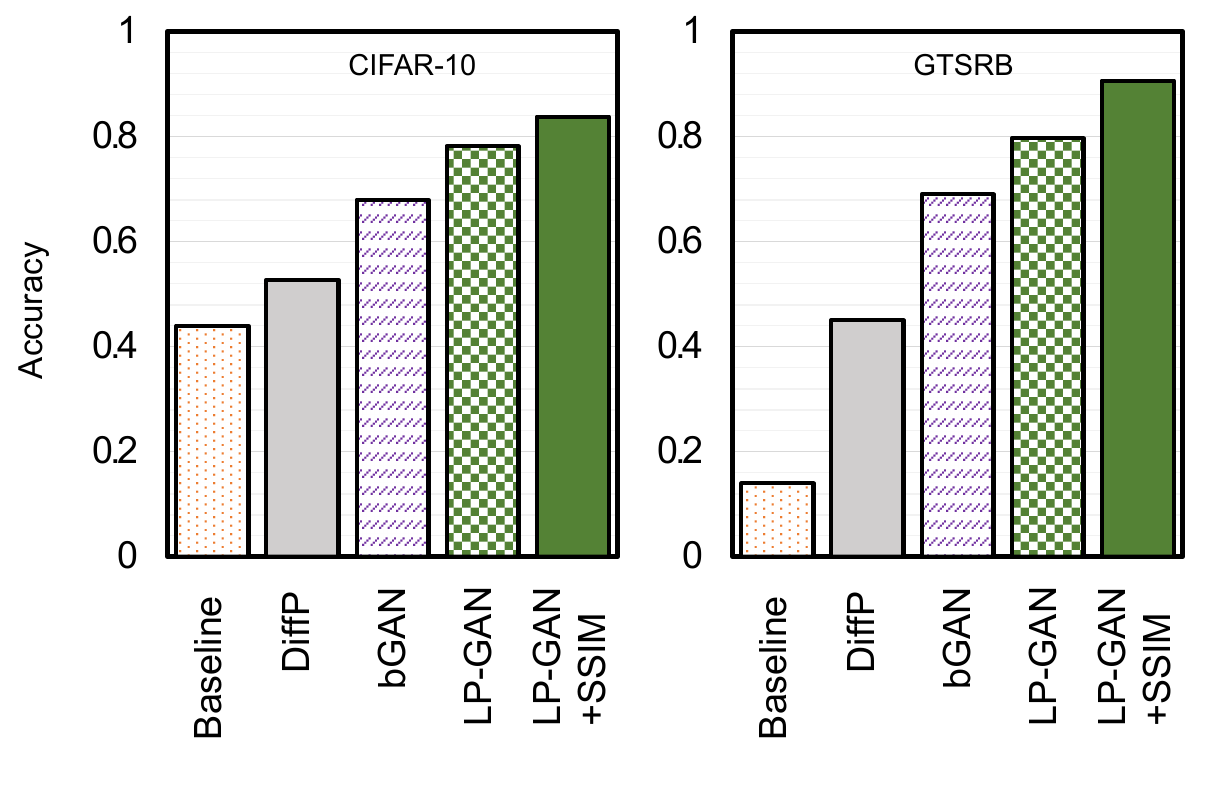}
         \caption{Gray-Box.}
         \label{fig:gray}
     \end{subfigure}
     \hfill
\begin{subfigure}[]{0.32\textwidth}
         \centering
         \includegraphics[width=\textwidth]{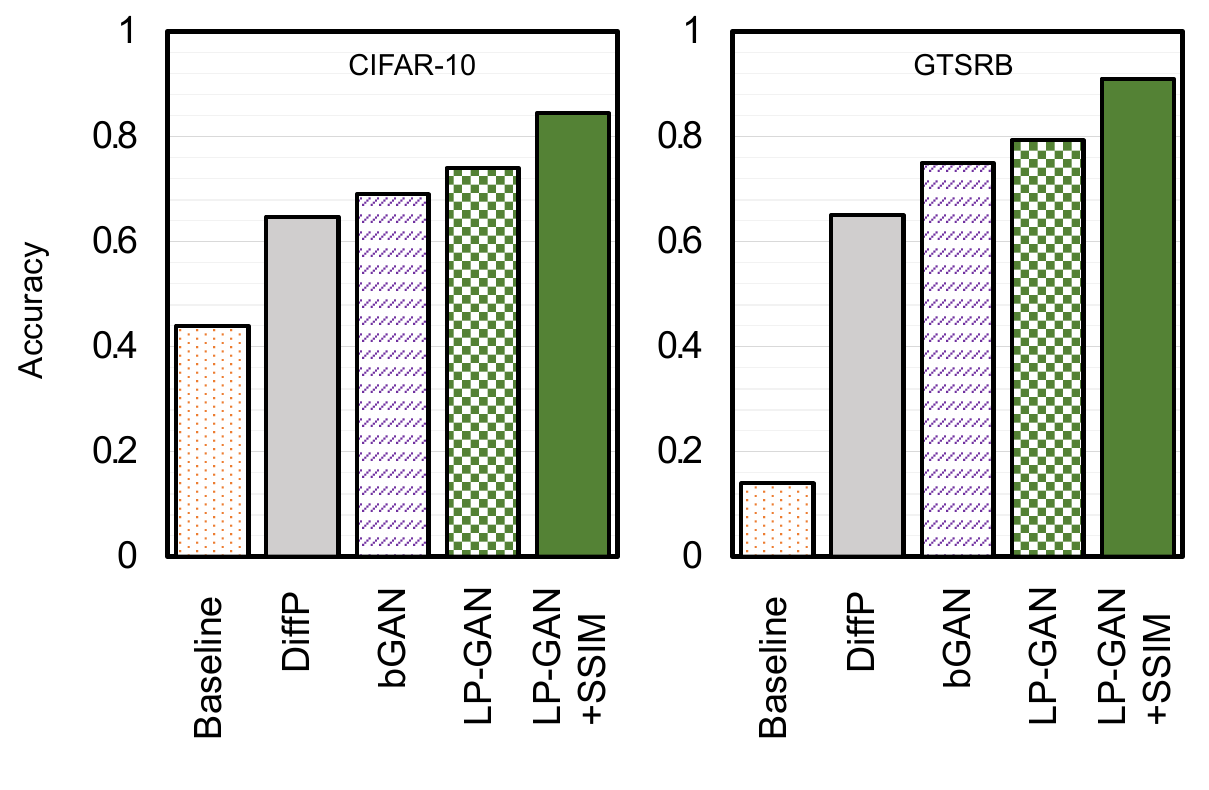}
         \caption{Black-Box.}
         \label{fig:black}
     \end{subfigure}
    \caption{Robustness for three different configurations using CIFAR-10 and GTSRB datasets. The results are for the baseline (no protection), Diffpure (DiffP), a baseline GAN (bGAN), and \textit{our method} (\sys) which includes the latency-aware diffusion model (LP-GAN) and the diffusion model with the accuracy-aware model (LP-GAN+SSIM).}
    \label{fig:rob}
\end{figure*}

To avoid out-of-memory issues, prior purification methods such as Diffpure calculate the gradients using an adjoint method, which has been reported to be sensitive to numerical errors \cite{nie2022diffusion, zhuang2020adaptive}. When compared to an attack using full gradients obtained from direct backpropagation, the adjoint method returns a higher robust accuracy, leading to an overestimation of the robustness of the model \cite{lee2023robust}. Therefore, in this paper, we utilize full gradients for all attacks.

\vspace{3pt}
\noindent \textbf{Datasets.} We use \textit{three} standard datasets. \textbf{\textit{CIFAR-10}} is a 10-class dataset comprising 50,000 training samples and 10,000 test samples. Each sample is a 32×32 RGB color image. The 10 classes include airplanes, cars, birds, cats, deer, dogs, frogs, horses, ships, and trucks. We use only the Test dataset to evaluate our accuracy and robustness, and to train our purifier we use only the Train dataset, but we do not use any label to train our purifier.

\textbf{\textit{GTSRB}}  (German Traffic Sign Recognition Benchmark) is a 43-class dataset with over 50,000 images ranging in size from 15×15 to 250×250 pixels.  We divide them into training, and test datasets with 45000 and 5000 samples, respectively. Also, we resize the pictures to 32x32 for training and evaluation.

\textbf{\textit{Tiny ImageNet}} is a 200-class dataset derived from ImageNet~\cite{deng2009imagenet}, containing 100,000 training images and 20,000 validation/test images, each resized to 64×64 pixels. We used 5,000 images from the validation set for our evaluation~\cite{le2015tiny}.

\vspace{3pt}
\noindent \textbf{Classifiers.} For CIFAR-10, we use Resnet-56 \cite{chenyaofo}, a publicly available state-of-the-art classifier with a top-1 accuracy of 94.37 percent on clean images. Resnet-56 has 855,770 parameters which takes up 3.3MB of memory in our embedded board (Jetson Orin Nano).
For GTSRB, we train a Resnet-20. It achieves 97.26 percent accuracy on clean images. It has 272,474 parameters and takes up 1.04MB. For Tiny ImageNet, we train a ResNet-18 model. It achieves 45\% accuracy on clean images. The model has approximately 11.2 million parameters and takes up 44.6MB. 

For black-box attacks, we use Resnet-50 as the \textit{shadow} model which will be used by the adversary to create the adversarial samples. 

% 92, w: 51, g: 75, b:92
% 93, w: 70, 
%latency: us 96 ms, baseline: 10, ddgan=260 
%classifier black box: resnet20, resnet 50 

\vspace{3pt}
\noindent \textbf{State-of-the-Art Models.} In addition to \sys, we also implement other models on Jetson and reported latency, accuracy, and robustness results. Specifically, we implement Diffpure~\cite{nie2022diffusion}.  We test performance with the purifying diffusion model using 10, 20, and 30 timesteps. The diffusion model has 35,746,307 parameters which uses 136.4MB of memory. Additionally, we compare \sys with a GAN-only design (i.e., when no diffusion is used) based on the method proposed in Defense-GAN~\cite{defensegan}. Lastly, we compare our method with a recently proposed technique, \textit{Moving Target Defense} (MTD)~\cite{song2019moving}. We assume that MTD has at least 10 base models (for voting) with early stopping. Depending on the classifier, the MTD's storage overhead changes. On average, for CIFAR-10, it takes about 35MB of Jetson's memory. For GTSRB, the storage is about 12MB. Detailed results for different metrics and models are provided in Section~\ref{sec:result}.

\section{Results} \label{sec:result}

\subsection{Latency, Accuracy, and Robustness}
\noindent \textbf{Robustness.} We report the robustness of our method against the attack model and datasets described in Section~\ref{sec:setup}. We report the results for the baseline (no protection), Diffpure (DiffP)~\cite{nie2022diffusion}, a GAN-only solution (i.e., without diffusion) based on Defense-GAN~\cite{defensegan}, which we refer to as baseline GAN (bGAN), and our method (\sys). For our model, we report two configurations: the latency-aware diffusion model (LP-GAN) and the diffusion model with the accuracy-aware model (LP-GAN+SSIM). Results are reported for two primary datasets (CIFAR-10 and GTSRB), with additional comparisons made on Tiny ImageNet.

Results are reported for black-, gray-, and white-box adversary models (see Section \ref{sec:setup}), as shown in Figure \ref{fig:rob}. 

Our evaluations demonstrate that \textit{\sys consistently achieves the highest robustness across different attack configurations}. While black-box configuration is evidently the most secure, \sys still manages to achieve notable robustness under white-box and gray-box attacks (especially for the GTSRB dataset). Comparing the two datasets, it is much harder to defend against white-box attacks (even with Diffpure) on CIFAR-10 than on GTSRB, indicating that internal information is more beneficial for fine-tuning APGD attacks on CIFAR-10. Furthermore, it should be noted that white-box attacks are considered less feasible in mobile systems due to the need for direct physical access, which is often not practical. In our view, gray-box attacks represent the most realistic scenario for an attack in a mobile system.

Looking more closely at grey-box results, \sys achieves $>$80\% robustness, on average, across datasets. Comparing LP-GAN and LP-GAN with SSIM, the results are improved by 5\% for CIFAR-10 and by 10\% for GTSRB, highlighting the usefulness of utilizing a similarity metric to further improve the generation accuracy.  

\begin{figure}
    \centering
    \begin{subfigure}{0.21\textwidth}
         \centering
         \includegraphics[width=\textwidth]{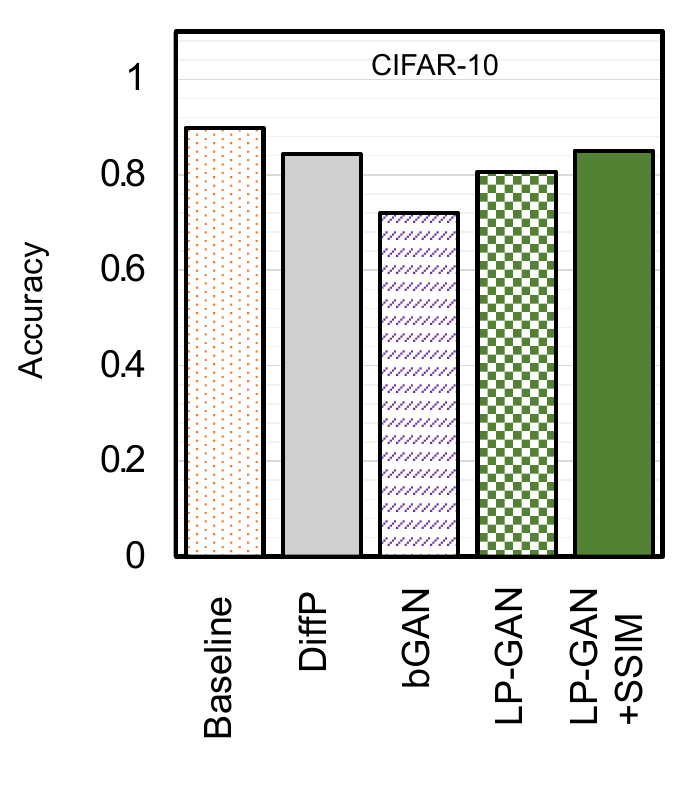}
         \caption{}
\end{subfigure}
\hspace{.05\textwidth}%
\begin{subfigure}{0.21\textwidth}
         \centering
         \includegraphics[width=\textwidth]{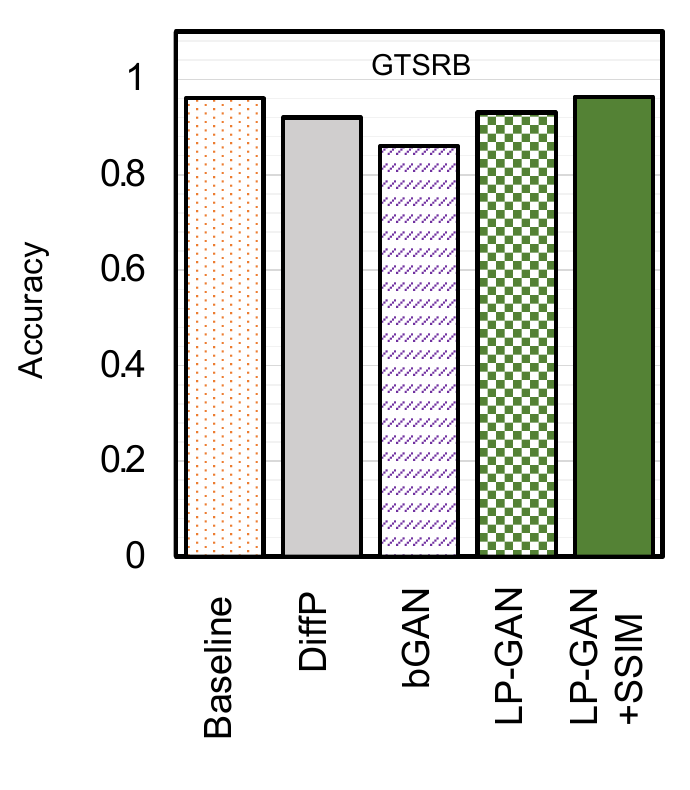}
         \caption{}
\end{subfigure}
    \caption{Clean accuracy (higher is better) for the downstream classification task using two different datasets.}
    \label{fig:acc}
\end{figure}

\vspace{3pt} \noindent\textbf{Accuracy.} Results for (clean) accuracy are reported in Figure~\ref{fig:acc}. Recall that we use two different classifiers (details in Section~\ref{sec:setup}) for the downstream classification task. It is important to analyze the impact of adversarial defense (e.g., purification, adversarial training, etc.) on the accuracy as retraining and/or adding and removing noise could potentially impact the clean accuracy.

As shown in Figure~\ref{fig:acc}, \sys incurs a minimal accuracy drop compared to the baseline. Specifically, for CIFAR-10 the drop for \sys is about 4\% when using the GAN+SSIM model. For GTSRB, the drop is almost zero ($<$.01\%). Compared to other methods, the usage of an accuracy-aware training model using SSIM loss and a generalized (two-step) discriminator/generator training method (see Section~\ref{sec:design}). 

\begin{figure}
    \centering
    \includegraphics[width=.8\columnwidth]{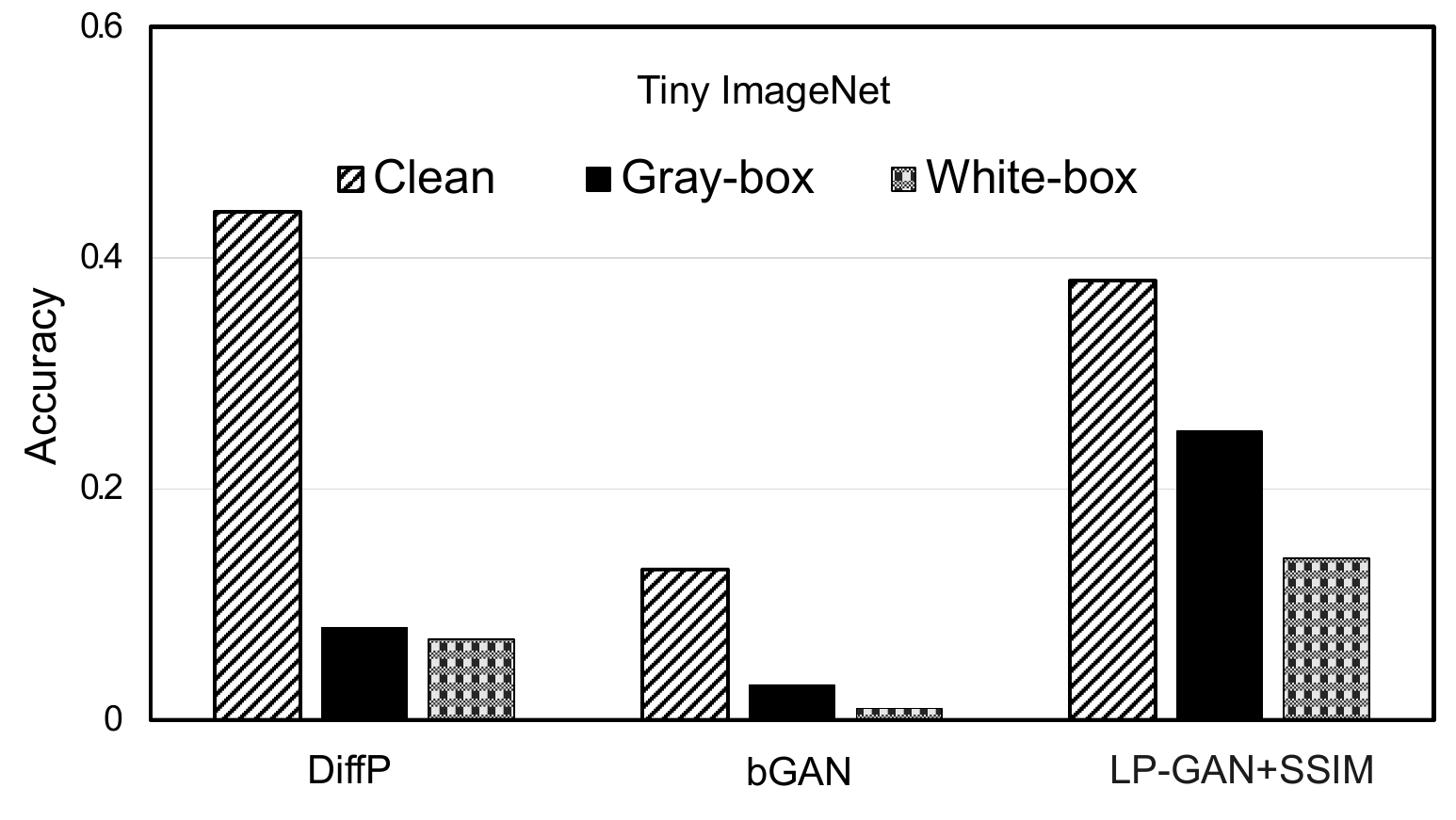}
    \caption{Comparison of accuracies across different models and attack scenarios on Tiny ImageNet.}
    \label{fig:accuracy_comparison}
\end{figure}

\vspace{3pt} 
\noindent\textbf{Tiny ImageNet Results.} While CIFAR-10 and GTSRB provide a solid basis for evaluating our method's robustness, we also tested our approach on the Tiny ImageNet dataset, a more challenging benchmark due to its numerous classes and reduced image size.  Results are shown in Figure \ref{fig:accuracy_comparison}. Our model outperforms both DiffPure and the Baseline GAN in gray-box and white-box scenarios. Also, note that the baseline accuracy in all cases is much lower than that of other datasets due to the complexity of the classification task. These results further highlight the robustness of our method, even in more complex datasets like Tiny ImageNet.

\begin{figure}
    \centering
    \begin{subfigure}{0.21\textwidth}
         \centering
         \includegraphics[width=\textwidth]{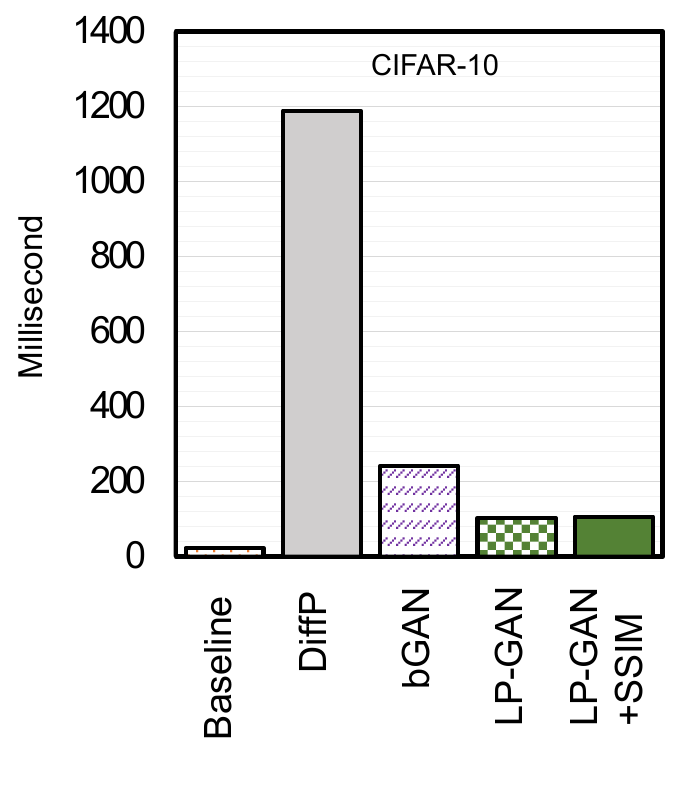}
         \caption{}
\end{subfigure}
\hfill
\begin{subfigure}{0.21\textwidth}
         \centering
         \includegraphics[width=\textwidth]{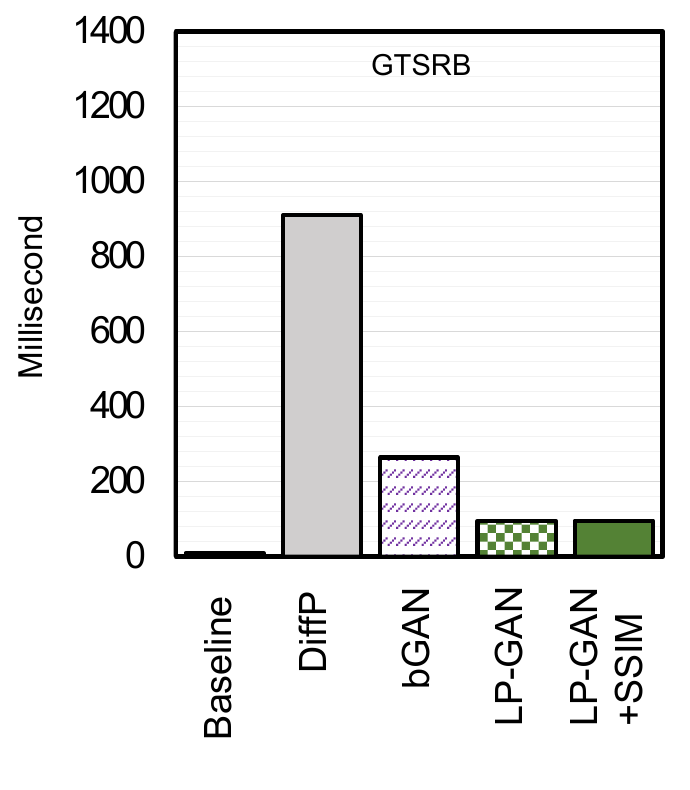}
         \caption{}
\end{subfigure}
    \caption{Latency for purification+classification for different methods (lower is better). y-axis shows the latency in milliseconds.}
    \label{fig:lat}
\end{figure}

\vspace{3pt} \noindent\textbf{Latency.} We compare the inference latency (purification and classification combined) for different methods (refer to Section~\ref{sec:imp} for details of the device and measurement methodology).  The results are displayed in Figure \ref{fig:lat} in milliseconds. 

Compared to the baseline (classification only), purification methods add between $>$50x to 4x. The reason for added latency is the extra processing needed during purification. Among different methods, \sys achieves the lowest. We improve DiffPure's latency by an order of magnitude since our method does not need multiple steps and each step is much simpler. Furthermore, compared to the baseline GAN method~\cite{defensegan}, \sys is more than twice faster mainly because the generation is one-shot as opposed to the multi-step generation needed in state-of-the-art GAN methods.  

Also, note that the latency for LP-GAN and LP-GAN+SSIM is the same since the generator architecture and steps stay similar for both methods. The difference is during the training phase which does not impact the inference latency.

\subsection{Comparison with Moving Target Defense (MTD) Method}
We further compare our results with the \textit{state-of-the-art non purification methods}. Specifically, we compare \sys (LP+SSIM) with MTD~\cite{song2019moving} in terms of accuracy, latency, and robustness. Instead of using purification, MTD leverages several base models (10-20 models) and develops a majority voting and detection method to eliminate adversarial attacks.

For accuracy and robustness, we use the results reported by the authors~\cite{song2019moving}. We use the gray-box attack configuration\footnote{MTD's threat model is neither black-box nor gray-box. To be fair, we assume the gray box in our setting which has lower robustness.}. For latency, we assume that MTD queries at least 10 base models (this is a favorable assumption) using the early stopping scheme (with $w=0.2$)~\cite{song2019moving}.

\begin{figure}
    \centering
    \begin{subfigure}{0.235\textwidth}
         \centering
         \includegraphics[width=\textwidth]{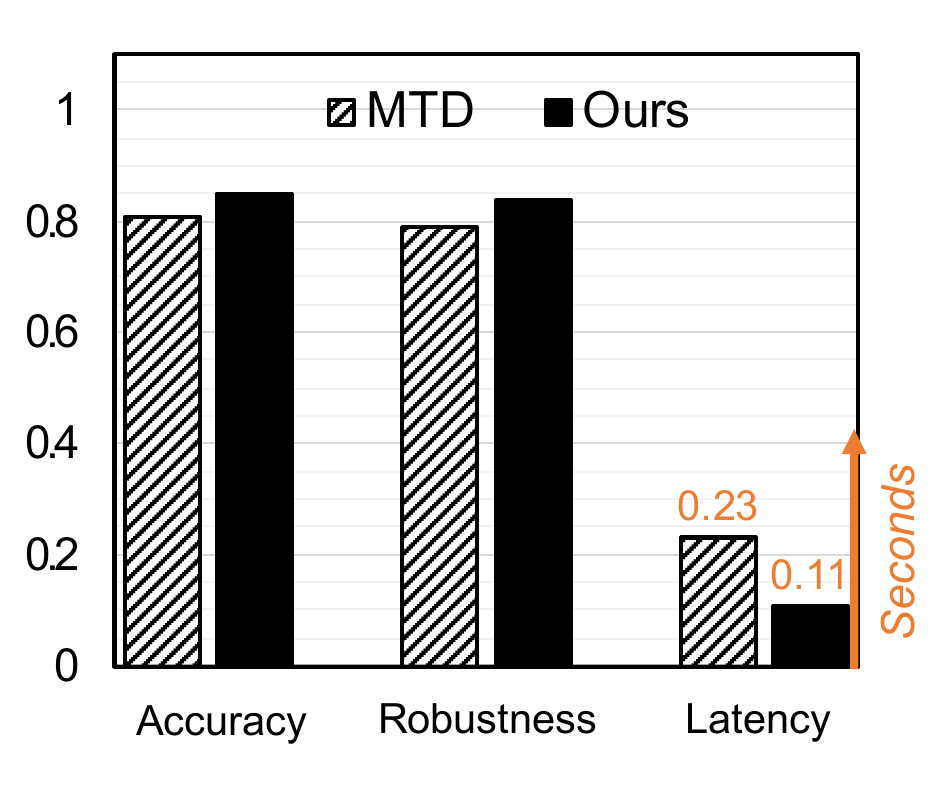}
         \caption{CIFAR-10}
\end{subfigure}
\hfill
\begin{subfigure}{0.235\textwidth}
         \centering
         \includegraphics[width=\textwidth]{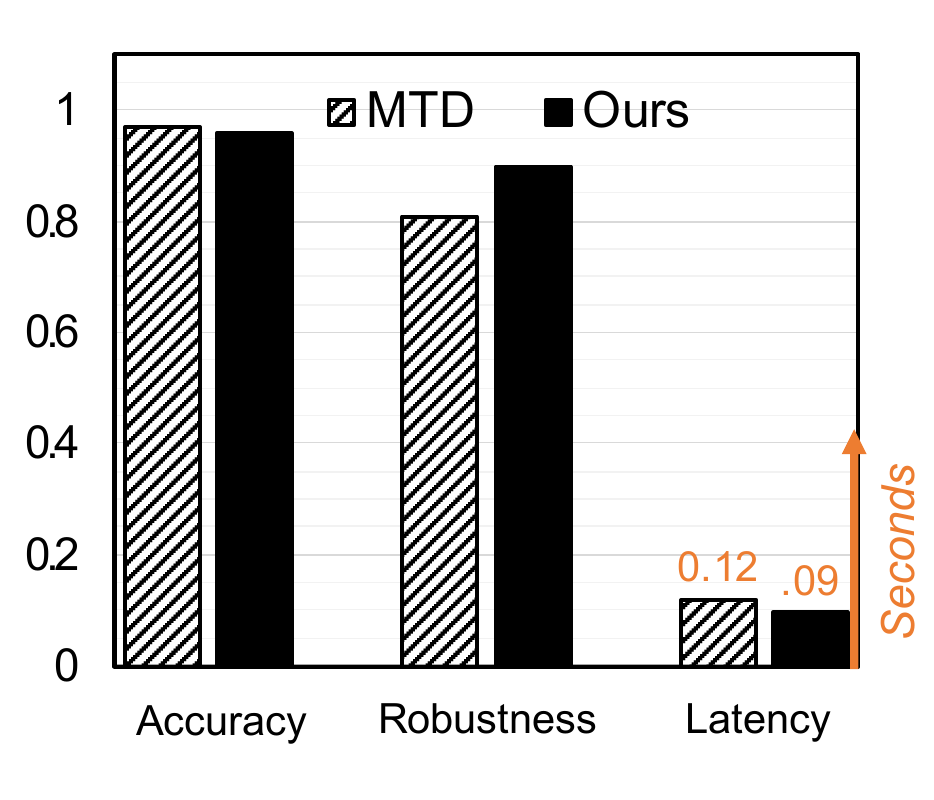}
         \caption{GTSRB}
\end{subfigure}
    \caption{Comparison between our method and state-of-the-art non-purification method, MTD~\cite{song2019moving}. Accuracy and robustness are normalized percentages (higher is better) while latency is in seconds (lower is better).}
    \label{fig:mtd}
\end{figure}

Results are shown in Figure \ref{fig:mtd}. For CIFAR-10, \sys achieves higher accuracy and robustness. The results are fairly comparable in general. The clear advantage of \sys, however, is in latency as it has more than 2x lower latency for the same classifier. The main reason for this is that \sys leverages a one-shot purifier while MTD needs several serialized queries from the classifier. The problem is exacerbated when using a more sophisticated classifier. 

Looking at the results for GTSRB in Figure~\ref{fig:mtd} (b), similar trends can be observed. Compared to the CIFAR-10 dataset, \sys achieves better robustness but similar accuracy (both around 96\%). Similar to CIFAR-10, \sys outperforms MTD in terms of latency. The difference is smaller since the classifier is faster (smaller). 

The storage overhead can be also compared. Based on the implementation results in Sections~\ref{sec:imp} and \ref{sec:setup}, \sys has about 174MB storage overhead. For MTD, the overhead depends on the classifier. For a smaller Resnet-18 classifier, this is about 12MB. For a larger Resnet-56 classifier, the storage overhead is about 35MB. Please note that while the storage overhead of \sys is higher, the overall storage requirement in \sys is still only 2\% of the total memory. In both cases, the storage requirement is negligible. 

%Lastly, the key benefit of \sys over MTD is that our method eliminates the need for retraining and/or modifying the classifier. %For large, complex, and/or proprietary classifiers, this is a significant advantage. 
Overall, the key takeaway is that \sys is more beneficial for larger classifiers when latency overhead is large. Furthermore, \sys is beneficial in scenarios where re-training is prohibitive or infeasible. Given the trend of using larger and more sophisticated models in autonomous mobile systems, we believe \sys provides an effective solution for current and future models. Additionally, in the following section, we'll demonstrate how the method can be further combined to improve the system's overall performance.

% some discussion on storage ... 

\subsection{\sys with Adversarial Training/MTD}
Since our proposed purification model is an orthogonal defense method to adversarial training and/or moving target defense, we can also combine our method with them by feeding the purified images from \sys (LP+SSIM) to the adversarially trained and/or MTD classifiers. 

\begin{figure}
    \centering
    \includegraphics[width=.65\columnwidth]{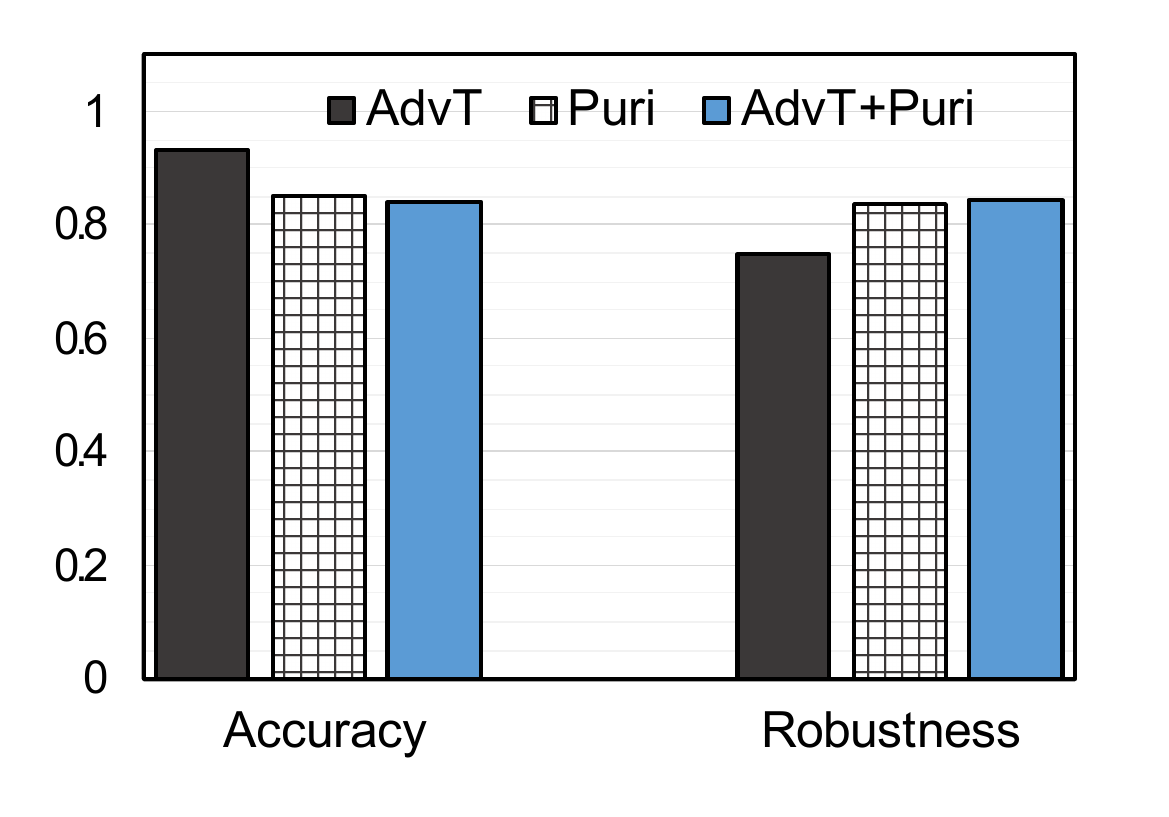}
    \caption{Accuracy and robustness results for combining adversarial training (AdvT) with our purification (Puri) model.}
    \label{fig:at}
\end{figure}

To experimentally evaluate this, Figure \ref{fig:at} shows that this combination (``AdvT + Puri'') can improve the robust accuracies against different attack models. The figure shows the accuracy and robustness under three different setups: \textit{1)} Training a robust classifier using adversarial training~\cite{croce2020reliable}. \textit{2)} Using \sys to purifier the image but using a regular classifier (not adversarially trained); and \textit{3)} Combining the robust classifier in (1) with our purifier (2). 

For robustness, we report the gray-box accuracy for all three cases. The results in Figure~\ref{fig:at} show that the overall robustness slightly improves when combining adversarial training (AdvT) with adversarial purification (Puri). This, however, comes at the cost of a slight reduction in accuracy as both AT and Puri cause degradation in image quality and overall classification accuracy due to adding perturbations to the model and/or input. Although not shown in this section, a similar approach could be used to combine Puri with MTD. Even further, Puri, AdvT, and MTD can be combined if desired.

The main takeaway from this experiment is that \sys proposes an orthogonal solution to existing non-purification solutions. When combined, the results could improve. However, the combination introduces new tradeoffs between accuracy, robustness, and training time. The designers can ultimately choose the right balance to achieve their goals. % Therefore, we can apply our method to the pre-existing adversarially trained classifiers for further improving the performance.

%\section{Discussions} \label{sec:discuss}
%\input{Sections/discussion}

\section{Further Analysis and Discussions}\label{sec:abst}
\subsection{Ablation Study}
As explained in Section~\ref{sec:design}, we utilize a two-step noise addition method during training. To highlight the importance of such a method, we initially experimented with a single-step noise addition process. However, we observed that this approach led to significant overfitting, particularly when using SSIM loss, as the generated image was directly compared to the original, uncorrupted image. \textit{Our results show more than 50\% reduction in robustness.} To address this issue, we introduced a two-step diffusion process. First, we generate a noisy image \(x_1\) from the original clean image. Rather than using \(x_1\) for direct training, we add an additional layer of noise to create a second noisy image \(x_2\). The model is then trained to denoise \(x_2\). After denoising \(x_2\), we calculate the SSIM loss between the original \(x_1\) and the \(x_1\) generated from the denoised version of \(x_2\). This step helps to increase generalization by ensuring that the model effectively handles various levels of noise, rather than merely replicating the clean images. This approach leads to better generalization and improved performance across different noise conditions.
\subsection{Comparison to Latent Diffusion Models}
In the context of adversarial image purification, we chose GAN-based networks over Latent Diffusion Models (LDMs) due to their superior effectiveness in preserving image accuracy. While LDMs are known for their speed, with a latency of around 300 ms- significantly faster than traditional Denoising Diffusion Probabilistic Models (DDPMs) \cite{rombach2022high, ho2020denoising}- this speed does not necessarily translate to better performance in purification tasks. Our GAN-based model, with a latency of 200 ms, not only offers comparable speed but also ensures higher purification effectiveness. The key advantage of GANs lies in their ability to operate directly in the pixel space, where adversarial noise is introduced, enabling more precise noise removal without the complications introduced by transforming images into a latent space, as LDMs do.

Transforming images into a latent space, as required by LDMs, can obscure the details necessary for effective noise removal and potentially shift decision boundaries, complicating the purification process and reducing robustness. In contrast, our GAN-based approach maintains the image in its original domain, ensuring that the purification process directly addresses the adversarial noise at the pixel level. This approach results in better preservation of the image's accuracy and robustness, making GAN networks a more suitable choice for adversarial image purification, where maintaining the integrity of the original image is crucial.

\subsection{Our Model Limitations}
Our model's limitations include the necessity to train our GAN separately for each dataset and using different noise levels to determine the optimal level for purification. Additionally, determining the appropriate ratio of SSIM loss to GAN loss presents another challenge. In contrast, DiffPure does not require retraining the diffusion model to identify the required noise level for purifying adversarial samples \cite{nie2022diffusion}. Consequently, finding these hyperparameters makes training our model more challenging compared to DiffPure, which utilizes a pre-trained diffusion model. 

\subsection{Exploring Potential Architectural Optimizations in StyleGAN Models}
Various architectural optimizations can be employed to enhance the efficiency and performance of our model. One area of interest is \textbf{pruning}, where reducing the model's redundancy by incrementally removing less impactful weights might decrease memory usage and improve computation speed. Other techniques such as structured and iterative pruning can be used to improve the latency without significantly impacting output quality \cite{han2015learning}. Additionally, \textbf{knowledge distillation} could be examined, where a smaller ``student'' model would be trained to replicate the ``teacher'' model. This approach would focus on matching both the output distribution and intermediate feature maps to potentially reduce the model size and accelerate inference times \cite{hinton2015distilling}.

\section{Related Work}\label{sec:relw}

\noindent\textbf{Adversarial Defenses.} First introduced by Madry \textit{et al.}, \textit{adversarial training} incorporates adversarial samples during training. Many improvements have been made to this technique, including the work done by He \textit{et al.}, which injects trainable Gaussian noise at each layer of the model to introduce randomness \cite{he2019parametric}. Gowal \textit{et al.} utilized generative models for data augmentation, with diffusion models resulting in the best robustness \cite{gowal2021improving}. 
A major drawback of adversarial training, however, is the need for retraining the classifier. 

\textit{Adversarial purification}, on the other hand, involves training a separate purification model that removes adversarial perturbations on input images before the downstream task. MagNet leverages auto-encoders to move adversarial examples closer to the manifold of legitimate examples \cite{meng2017magnet}, and Samangouei \textit{et al.} proposes using GANs as the purification model and uses GD minimization to yield higher robustness \cite{defensegan}. More recently, Song \textit{et al.} proposed PixelDefend, which relies on PixelCNN, an autoregressive generative model \cite{song2018pixeldefend}, and Yoon \textit{et al.} used score-based generative models to purify adversarial samples \cite{yoon2021adversarial}. Diffpure was the first attempt to use a diffusion process followed by a backward denoising process for purification \cite{nie2022diffusion}. 

As extensively discussed in this paper, \sys improves the prior work in two major directions. Compared to GAN-only based methods, \sys significantly improves robustness and accuracy while also slightly improving latency as it uses one-step forward and backward processes. Compared to diffusion-based models, \sys significantly improves latency while also achieving better robustness. 

\vspace{3pt}
\noindent \textbf{Attacks to Physical Autonomous Mobile Systems.} Also relevant to this work are methods that target physical sensors (e.g., camera, LIDAR, mmWave, RF, etc.) to attack autonomous mobile systems. Some attacks target the availability and/or integrity of the sensor by manipulating the environment and/or the operation of the sensors \cite{zhu2021adversarial}. 

More relevant to this work are methods that create physical adversarial samples to impact the machine learning-based classification task \cite{song2018physical, lovisotto2021slap, kong2020physgan}. These attacks can further be categorized as physical adversarial attacks \cite{brown2017adversarial} and physical sensor attacks \cite{zhu2023tilemask}. In both categories, there is a rich literature for creating adversarial patches \cite{liu2019perceptual}. They commonly use similar methods (e.g., FGSM, PGD, etc.) to those considered in this paper to generate adversarial samples. Depending on the physical modalities and threat model, various methods could be used to conduct the attack (e.g., shooting a laser, jamming the signal, causing interference, 3D printing the object, etc \cite{zhu2023tpatch}.)

Compared to these methods, we consider a strong and comprehensive threat model ranging from fully white-box (complete knowledge) to black-box. As mentioned in the paper, we believe that gray-box attacks are the most reasonable assumption in our setting.

\begin{table}[t]
    \centering
    \small
    \caption{Comparing different methods for visual analysis on latency, accuracy, robustness, and retraining. For each property, a filled circle means better. (AT = Adversarial Training, DG = Defense-GAN, DP = DiffPure, MTD = Moving Target Defense).}
    \begin{tabular}{c|c|c|c|c|c|}
        \cline{2-6}
        &  AT~\cite{croce2020reliable}& DG~\cite{defensegan}&  DP~\cite{nie2022diffusion}&  MTD~\cite{song2019moving}&  Ours\\
        \hline Latency& \fullcirc &  \halfcirc & \emptycirc &  \halfcirc& \fullcirc \\
        \hline Accuracy& \fullcirc& \halfcirc& \fullcirc & \fullcirc& \fullcirc \\
        \hline Robustness& \halfcirc& \halfcirc & \halfcirc & \fullcirc& \fullcirc \\
        \hline Retraining& \emptycirc & \fullcirc& \fullcirc& \emptycirc & \fullcirc \\
        % \hline Storage& \fullcirc & \halfcirc& \halfcirc&\emptycirc& \halfcirc\\
        \hline
    \end{tabular}
    \label{tab:compare}
\end{table}

\vspace{3pt}
\noindent\textbf{Comparison.} We conclude this section by summarizing different solutions. In Table \ref{tab:compare}, we compare adversarial training, defense-GAN, DiffPure, and MTD on the metrics of latency, accuracy, robustness, and retraining. 

\section{Conclusions} \label{sec:conc}
This paper addressed the critical challenge of defending autonomous mobile systems against adversarial machine learning attacks, which pose significant threats to their reliability and safety. While previous countermeasures have shown promise, they often require invasive modifications to classifiers or incur high computational costs, making them unsuitable for resource-constrained mobile devices.

To overcome these limitations, we introduced an innovative purification approach that leverages a GAN framework. Our method achieves a remarkable balance between classification accuracy, adversarial robustness, and computational efficiency, making it well-suited for real-world applications where speed and resource constraints are paramount. Our key contribution was the design and implementation of a new one-shot generator that leverages a two-step forward diffusion process during its training. The method was further improved by introducing a similarity-based loss function for training the generator. 

Through extensive experimentation and evaluation, we demonstrated the superiority of our approach over existing purification methods, achieving significant improvements in both latency and performance across various attack scenarios. Our method represents a notable advancement in the field of adversarial image purification, offering a scalable and effective solution for safeguarding autonomous mobile systems in practical settings.

\section*{Acknowledgement}
 We thank our shepherd for their guidance. This work has been supported, in part, by CISCO Systems Inc., NSF grants CNS-2211301,  CNS-2303115, and CNS-2312089,  and a gift received from Accenture. The views and findings in this paper are those of the authors and do not necessarily reflect the views of Cisco, NSF, and Accenture.

% \begin{acks}
% \end{acks}

\bibliographystyle{ACM-Reference-Format}

% %%
% %% If your work has an appendix, this is the place to put it.
% \appendix

% \section{Discussions}
% \input{Sections/discussion}

% \section{GAN Structure}
% The internal structure of our GAN is shown in Figure~\ref{fig:gan}. 
% \begin{figure}[h]
%     \centering
%     \includegraphics[width=1\columnwidth]{Figures/gan.pdf}
%     \caption{The internal architecture of our GAN model used for denoising.} % numbers should be enlarged}
%     \label{fig:gan}
% \end{figure}
% %\section{Research Methods}

\end{document}